\documentclass[10pt, conference]{IEEEtran}

\usepackage[square,numbers,sort&compress]{natbib}
\usepackage{tabularx, color, colortbl}
\usepackage{graphicx, subfigure}
\usepackage{amsmath, amssymb}
\usepackage[vlined, ruled, linesnumbered]{algorithm2e}
\usepackage{listings}
\usepackage[noblocks]{authblk}
\usepackage{mathtools}
\usepackage{todonotes}
\usepackage{framed}
\usepackage{subfig}
\usepackage{soul}
\usepackage{verbatim}
\usepackage{url}
\usepackage{booktabs}
\usepackage{flushend}
\usepackage{comment}
\usepackage{multirow}
\usepackage{subfiles}
\usepackage{booktabs}
\usepackage{caption}
\usepackage{subcaption}

\usepackage{pifont}   
\usepackage{float}    
\usepackage{xcolor}   

\newcommand{\cmark}{\textcolor{blue}{\ding{51}}} 
\newcommand{\xmark}{\textcolor{red}{\ding{55}}}   

\newcommand*\circled[1]{\tikz[baseline=(char.base)]{
\node[shape=circle,fill,inner sep=1pt] (char) {\textcolor{white}{#1}};}}

\usepackage[letterpaper, margin=0.75in]{geometry}

\usepackage{cleveref}
\crefname{section}{§}{§§}
\Crefname{section}{§}{§§}

\def\BibTeX{{\rm B\kern-.05em{\sc i\kern-.025em b}\kern-.08em
    T\kern-.1667em\lower.7ex\hbox{E}\kern-.125emX}}

\begin{document}



\title{DIAL: Decentralized I/O AutoTuning via Learned Client-side Local Metrics for Parallel File System}

\author{
    Md Hasanur Rashid\IEEEauthorrefmark{1}, 
    Xinyi Li\IEEEauthorrefmark{2},  
    Youbiao He\IEEEauthorrefmark{2}, 
    Forrest Sheng Bao\IEEEauthorrefmark{2},
    Dong Dai\IEEEauthorrefmark{1} \\
    
    \IEEEauthorrefmark{1}Department of Computer and Information Sciences, University of Delaware, Newark, US \\
    \IEEEauthorrefmark{2}Department of Electrical and Computer Engineering, Iowa State University, Ames, US \\
    
    Email: \{mrashid, dai\}@udel.edu, \{xinyili, yh54, fsb\}@iastate.edu
}

\maketitle 
\IEEEpubidadjcol

\begin{abstract}
Enabling efficient, high-performance data access in parallel file systems (PFS) is critical for today's high-performance computing systems.
PFS client-side I/O heavily impacts the final I/O performance delivered to individual applications and the entire system.
Autotuning the key client-side I/O behaviors has been extensively studied and shows promising results. However, existing work has heavily relied on extensive number of global runtime metrics to monitor and accurate modeling of applications' I/O patterns. Such heavy overheads significantly limit the ability to enable fine-grained, dynamic tuning in practical systems. In this study, we propose DIAL (\underline{D}ecentralized \underline{I}/O \underline{A}utoTuning via \underline{L}earned Client-side Local Metrics) which takes a drastically different approach. Instead of trying to extract the global I/O patterns of applications, DIAL takes a decentralized approach, treating each I/O client as an independent unit and tuning configurations using only its locally observable metrics. With the help of machine learning models, DIAL enables multiple tunable units to make independent but collective decisions, reacting to what is happening in the global storage systems in a timely manner and achieving better I/O performance globally for the application.
\end{abstract}
	
\pagestyle{plain} 
  
\section{Introduction}
\label{sec:intro}
High-performance computing (HPC) is a cornerstone for modern science.
HPC's transformative power lies not just in computational speed but also critically in its I/O performance.
However, achieving optimal I/O performance is not an easy task due to the increasingly complex HPC storage architecture, which typically involves multiple shared parallel file system (PFS) components and long I/O paths~\cite{rashid2023iopathtune}. 

Automatic tuning is considered a promising solution. Existing efforts naturally take the approach of \textit{understanding the global I/O patterns of applications and configuring the I/Os accordingly}. Behzad et al. proposed a pattern-driven I/O tuning approach, which trains I/O prediction models tailored to specific I/O patterns~\cite{behzad2019optimizing}. With I/O pattern modeling, different strategies to search for the best configurations have been studied~\cite{cao2018towards, lyu2020sapphire}. Recent studies, such as CAPES~\cite{li2017capes}, incorporated deep reinforcement learning (DRL) to provide end-to-end I/O parameters tuning. In these methods, the I/O patterns are also implicitly modeled by the RL agents.

However, such strategies face a critical challenge. Fundamentally, detecting the I/O patterns of large-scale HPC applications is hard, not to mention effectively using the detected patterns to navigate through the large parameters' value space. The accuracy of I/O pattern detection can be compromised by various factors, such as inadequate observed metrics, I/O interference from other applications, and I/O noises on storage devices~\cite{isakov2022taxonomy, egersdoerfer2024understanding}. Models that seem to perform well often rely heavily on a large number of runtime metrics, and many of them need to be global~\cite{li2017capes}, which introduce high overhead for both collecting the metrics and training the models.

To this end, we propose a novel auto-tuning framework \textbf{DIAL} (\textbf{D}ecentralized \textbf{I}/O \textbf{A}utotuning via \textbf{L}earned Client-side Local Metrics), designed to adjust tunable I/O parameters of PFS I/O clients, relying solely on learned client-side local metrics. Running on each I/O client, DIAL framework periodically probes the local PFS, extracts low-level metrics from raw system statistics, then leverages the trained machine learning model to decide if better configurations can be applied to improve its I/O performance. If yes, it selects the best configuration and immediately applies it to the system. Such a procedure continues periodically in a user-defined interval so that the framework is able to catch the latest changes in both applications' I/O behaviors and the global storage system statuses. 
Our contributions are twofold:
\begin{itemize}
    \item We developed a novel autotuning framework that is capable of online adaptive parameter tuning independently and relies solely on local metrics.
    \item We identified key learned low-level metrics derived from clients that effectively characterize the status of clients' I/O activity and the global storage systems.
\end{itemize}

The rest of the paper is organized as follows: In \S\ref{sec:background}, we discuss relevant backgrounds of the widely used Lustre file system and motivations behind some of DIAL's design choices. In \S\ref{sec:design}, we present the architecture of DIAL in detail. We present the extensive experimental results in \S\ref{sec:eval}. We highlight the relevant works to our study in \S\ref{sec:related}, conclude the paper and discuss the future work in \S\ref{sec:conclude}.

\section{Lustre and the Key Idea}
\label{sec:background}

\subsection{Lustre Overall Architecture}
\begin{figure}[htpb]
    \centering
    \includegraphics[width=0.32\textwidth]{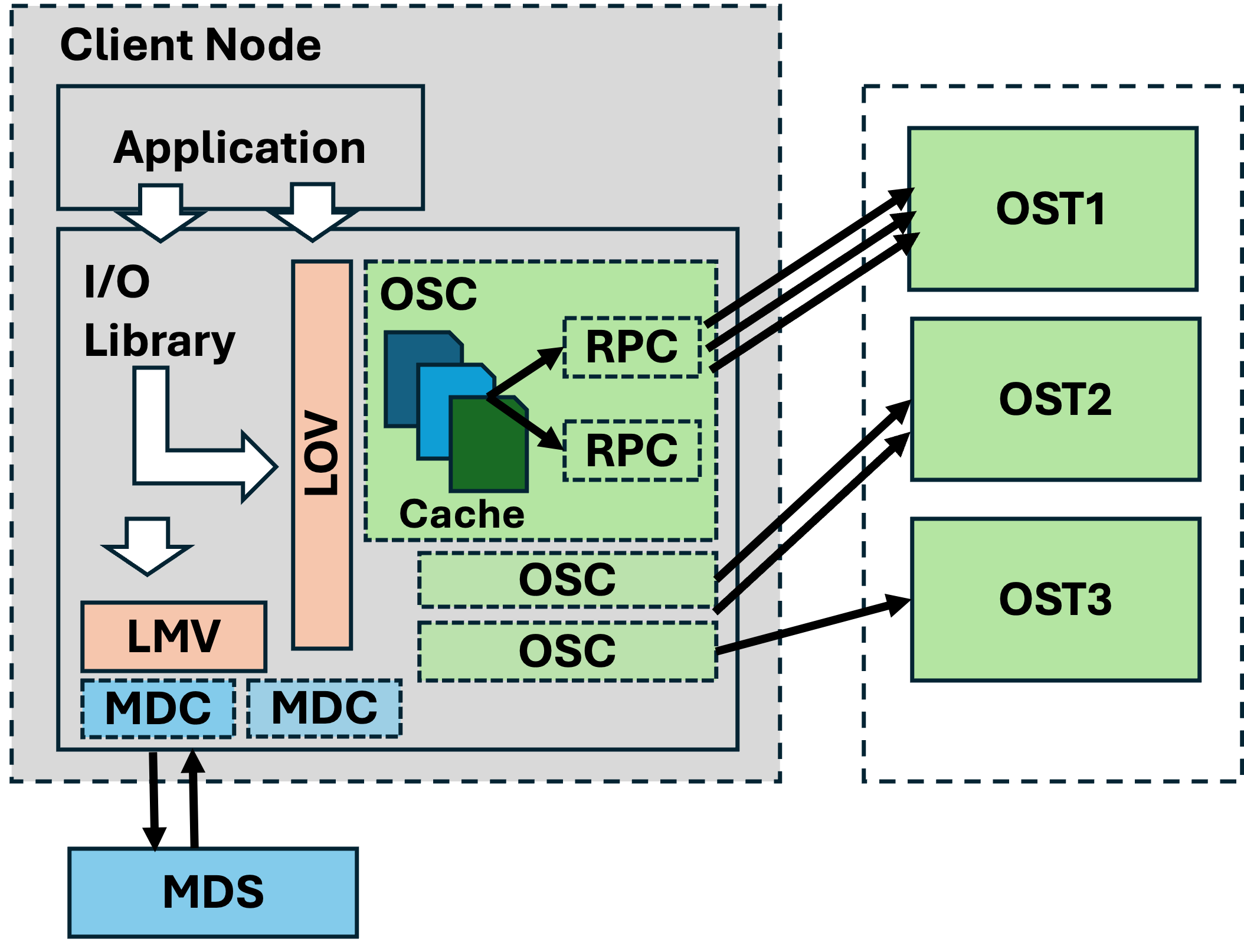}
    \caption{The detailed I/O path of Lustre.}
    \label{lustre-iopath}
    \vspace{-1em}
\end{figure}

A typical Lustre~\cite{braam2019lustre} cluster includes one management server (MGS), one or more metadata servers (MDSs), and many object storage servers (OSSs). The management server (MGS) is very lightweight and normally deployed on one of the MDSs. Lustre employs remote procedure calls (RPCs) to enable communication between clients and servers.
The Lustre client Virtual File System (VFS) interface is called Lustre Lite (LLITE), which is the bridge between the Linux kernel and the underlying Lustre infrastructure represented by the Lustre Object Volume (LOV) or Lustre Metadata Volume (LMV). 
For data accesses, the I/O requests will be handled by the LOV component, which establishes multiple Object Storage Client (OSC) interfaces, each corresponding to one Object Storage Target (OST) on OSS. Each OSC interface is responsible for formulating RPCs and managing RPC communications directed to a specific OST. To improve I/O performance, each OSC interface maintains its own local buffer to cache `dirty' data that has been recently read or written.  Similarly, the Lustre Metadata Volume (LMV) component manages metadata accesses to metadata targets (MDTs) on MDSs via the Metadata Client (MDC) interface. 
Figure~\ref{lustre-iopath} further shows how I/O requests are formulated from the client side in Lustre.

\subsection{Lustre Parameters for Autotuning}
Based on how Lustre clients handle I/O requests, we consider two parameters, `\textit{RPC Window Size}' and `\textit{RPCs in Flight}', to be ideal candidates to validate our idea. 
Firstly, their impacts on I/O performance are entangled with applications' I/O patterns and reflect the common property of many parameters. 
For instance, although maximizing the RPC window size may make each RPC channel work efficiently by sending a large chunk of data each time, facing a random small I/O access pattern may underutilize multiple RPC channels because not enough RPCs are formulated. The OSC will hold the transfer of RPCs, potentially idling RPC channels. 
%
For this reason, validating our idea over such parameters and their complicated relationships provides a solid foundation for other configurations in the system. 

Secondly, these two parameters exist in every client OSC interface, can be dynamically changed during runtime, and take effect in almost real-time, making them ideal for validating our idea. Many parameters in Lustre, such as stripe count and stripe size, can not take effect in real-time after values change. Making the entire parallel file systems fully tunable at runtime is our long-term research goal, but it involves different sets of challenges in software engineering, such as centralizing controls to each parameter or enabling immediate impacts of parameter changes. We consider them as future work, especially after validating our key idea here.

\begin{figure} [htpb]
    \centering
    \includegraphics[width=0.45\textwidth]{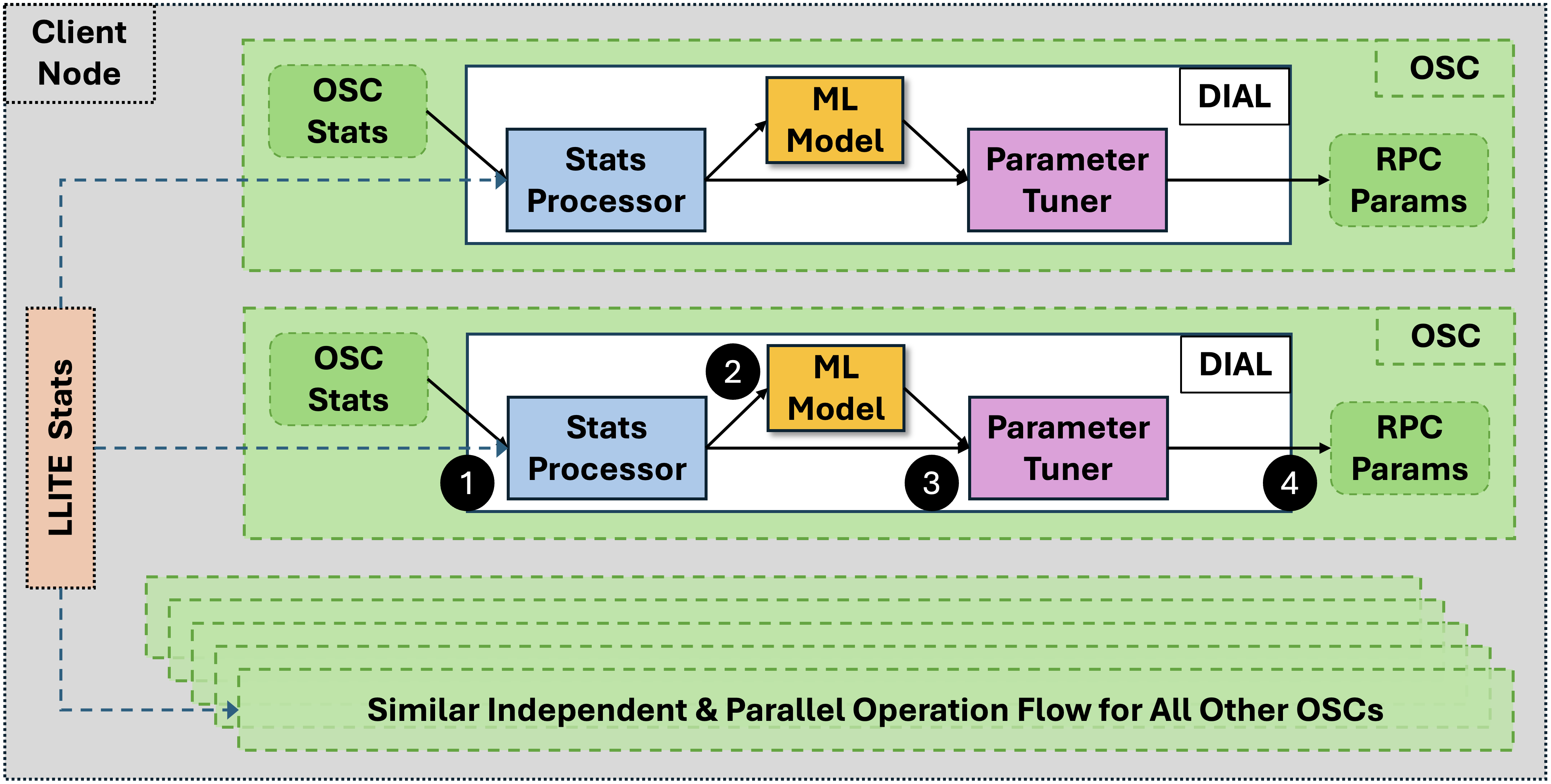}
    \caption{Architecture of the DIAL framework}
    \label{fig:iotuning}
    \vspace{-1em}
\end{figure}

\section{Design and Implementation} 
\label{sec:design}
\subsection{Overall Design}
In this study, we scrutinized the key idea by implementing a prototype based on the Lustre file system. Specifically, the proposed autotuning framework, DIAL will operate autonomously on each Lustre client, drawing solely on local file system statistics from the PFS client side. As illustrated in Figure~\ref{fig:iotuning}, the DIAL's architecture begins with the system stats collector and preprocessor component (\circled{1}), which probes the raw statistics from the file system at user-defined intervals (e.g., $1$ second). These statistics are then preprocessed to extract designed metrics. We then create the snapshots incorporating these metrics for the OSC interface. The processed OSC snapshots (\circled{2}) are fed into the local machine-learning model component, which generates a probability distribution across the configuration space. This distribution assesses the likelihood of each configuration achieving performance improvement if applied. Armed with the probability distribution and the OSC snapshots for each interface, the parameter tuner (\circled{3}) component determines the adjustments needed for the current configuration of each OSC interface and implements these changes (\circled{4}). This whole process repeats at each interval and continues throughout the duration of DIAL's operation on the client, ensuring continuous optimization.

\subsection{Machine Learning Model Design}
\label{subsec:ML_model}

We utilize machine learning models to capture the characteristics of our designed low-level metrics, aiming to comprehend \textit{how the current values of client metrics might influence I/O and how should a particular configuration would impact the performance accordingly}.

To define the machine learning model, let $\mathbf{s}_t$ represent a vector of system metrics at timestep $t$. A short  history of system performance of $k$ timesteps is thus $\mathbf{H}_t= [\mathbf{s}_{t-k}, \mathbf{s}_{t-k+1}, \cdots, \mathbf{s}_t]$. We further denote the tunable parameters of the system at timestep $t$ as $\mathbf{\theta}_t$, which is also a vector. The goal of a machine learning model here is to find a function $f$ that predicts the likelihood that the setting $\theta_t$ will result in a performance improvement over a threshold $\epsilon$ (e.g., 15\%) in the next timestep, thus $f(\theta_t, \mathbf{H}_t) = \mathbb{E}[ s_{t+1}/ s_{t} > 1+\epsilon]$ where $\mathbb{E}$ is expectation. In our study, GBDTs is chosen as the ML architecture with $k=1$.

We train separate models for read and write operations. This differentiation stems from observing the distinct approaches the Lustre file system employs in handling these operations. Write operations are subject to intricate RPC formation rules influenced by factors such as the distributed locking mechanism, grant utilization, and the OSC's cache utilization constraints, which are more complex than those for read operations. We design metrics specific to the operation type to capture operation-specific characteristics more precisely. By having separate models, we can more accurately predict the likelihood of I/O performance improvement based on our designed low-level system metrics and the specific configurations we aim to adjust. In the next section, we discuss how integrating metrics derived from the system and insights generated from the ML model enables us to fine-tune our tuning actions on the parameters.

\begin{algorithm}[htpb]
\caption{Parameter Tuning Algorithm}
\label{alg:calc_score}

\KwData{configuration space $\Theta$, a short history of the system $\mathbf{H}_t = [s_{t-k}, s_{t-k+1}, \cdots, s_t]$, operation type $o$, current configuration $\theta_t=\{\theta_t^1, \theta_t^2\}$, probability threshold $\tau$, weights $\alpha,\beta$}
\KwResult{Optimized configuration $\theta^*$}

$S=\emptyset$\;

\ForEach{$\theta \in \Theta$}{
    $p \gets f(\theta, \mathbf{H}_t)$\;
    \If{$p > \tau$}{
        $S \gets S \cup \{\theta\}$\;
    }
}

Normalize configurations in $S$ using MinMax normalization\;

\eIf{$o == \text{`Write'}$}{
    $\theta^* \gets \arg\max_{\theta \in S} \texttt{WriteScore}(\theta, \mathbf{H}_t)$\;
}{
    $\theta^* \gets \arg\max_{\theta \in S} \texttt{ReadScore}(\theta, \mathbf{H}_t)$\;
}

\SetKwFunction{FWriteScore}{WriteScore}

\SetKwProg{Fn}{Function}{:}{\KwRet}
  \Fn{\FWriteScore{$\theta, \mathbf{H}_t$}}{
    \Return 
        $f(\theta, \mathbf{H}_t) \cdot (1+ \beta \texttt{sum}(\theta))$\;
  }

\SetKwFunction{FReadScore}{ReadScore}

\SetKwProg{Fn}{Function}{:}{\KwRet}
  \Fn{\FReadScore{$\theta, \mathbf{H}_t$}}{
    \Return 
        ($f(\theta, \mathbf{H}_t) \cdot (1 + \alpha \theta^1)) + \theta^2$\;
  }

\end{algorithm}

\subsection{Parameter Tuning Strategy}
We adopt separate tuning strategies for \texttt{read} and \texttt{write} operations due to the distinct ways in which Lustre handles each operation, as detailed in the section \S~\ref{subsec:ML_model}. DIAL selects the appropriate trained ML model based on the volume of observed I/O operations (\texttt{Data Transfer Volume}) during the observation period. The ML model assigns a probability to each input configuration, providing an approximation of which configurations are likely to enhance I/O performance.
In this study, we introduce a \textit{Conditional Score Greedy} approach that first only considers configuration with probability exceeding the threshold, then calculates a \textit{priority score} based on the probability itself and a generic trend of the given configurations. 

Algorithm~\ref{alg:calc_score} outlines the tuning strategy. For each configuration $\theta$, the machine learning model will first calculate a probability $p$ of improving the performance by at least 15\%. If the probability is over a preset threshold $\tau$, the corresponding $\theta$ is added to the set $S$ for further consideration. We set $\tau$ as $0.8$ in our experiment. The optimal $\theta$ cannot be greedily defined as $\arg\max_\theta f(\theta, \mathbf{H}_t)$ as it leads to greedy tuning, which prefers high probability, safe configurations. Thus, we add regularization terms to them (lines 12 and 14).
Such a regularization is introduced as part of the \textit{Conditional Score Greedy} approach to fast converge the explorations. 

Specifically, we prioritize the larger values of the two selected parameters (`\textit{RPC Window Size}' and `\textit{RPCs in Flight}') when several configurations can all improve the performance.  This is because, at a high level, a larger RPC size would utilize network channels better and more RPCs in flight would transfer more data in parallel. Hence, they have a higher chance of being optimal configurations. Of course, higher values alone certainly would lead to extremely bad performance, which necessitates this study and our machine learning models. The algorithm contains two parameters $(\alpha,\beta)$ to help us adjust what level of importance we assign to the $\theta$ values. 

\section{Evaluation}
\label{sec:eval} 
\subsection{Cluster Setup \& Data Collection}
All evaluations were conducted on the open CloudLab platform, allowing for reproducible and fair comparisons.
We have used ten CloudLab c6525-25g machines, as detailed in Table~\ref{tab:hardware_spec}, to set up a prototype cluster.
We deployed Lustre version 2.15.5 as the PFS, with one dual-purpose machine serving as both the MGS and MDS, and four OSSes. Each OSS was configured with two OSTs, mapped to two SSD partitions. The remaining five machines were configured as Lustre clients.


\begin{table}[htpb]
\footnotesize
\centering
\caption{Hardware Specification for c6525-25g Node}
\begin{tabular}{ll}
\toprule
\textbf{Component} & \textbf{Specification} \\ 
\midrule
CPU   & 16-core AMD 7302P at 3.00GHz \\ 
\midrule
RAM   & 128GB ECC Memory (8x 16 GB 3200MT/s RDIMMs) \\ 
\midrule
Disk  & Two 480 GB 6G SATA SSD \\ 
\midrule
NIC   & Two dual-port 25Gb GB NIC (PCIe v4.0) \\ 
\bottomrule
\end{tabular}
\label{tab:hardware_spec}
\end{table}

The ML models were trained using the simplest workloads in Filebench benchmark, which included single stream I/O patterns (a single process accessing a large file on a single OST) with varying access patterns (sequential or random) and request sizes (small: 8 KB, medium: 1 MB, large: 16 MB).To collect training data, we simulated both \texttt{Read} and \texttt{Write} operations using specified I/O patterns. Each pattern was executed for 300 seconds and repeated 30 times to generate a sufficient number of samples. During these simulations, we probed the system at half-second intervals. The offline training dataset consists of 100,730 read-only and 98,078 write-only non-zero samples.

\subsection{Tuning Real-world Applications}
\label{sec:real_execution}

We tested the framework's effectiveness in tuning traditional HPC applications. We employed the H5bench benchmark, which includes I/O kernels developed based on the I/O patterns of a particle physics simulation (the \texttt{VPIC-IO} kernel for writing data) and a big data clustering algorithm (the \texttt{BDCATS-IO} kernel for reading data previously written by \texttt{VPIC-IO}). The sync write and read operations were tested under various conditions. For writing, we chose contiguous memory and file patterns, while for reading, we evaluated partial and strided read operations. The results in Table~\ref{tab:hpc_app} showed performance on par with the optimal configuration, corroborating our previous findings that, DIAL can effectively deliver near-optimal performance.

\setlength{\belowdisplayskip}{-10pt} 
\begin{table}[htbp]
\centering
\caption{HPC Scientific Application Executions}
\footnotesize
\begin{tabular}{ccc}

\toprule
\textbf{Scientific} & \textbf{Optimal} & \textbf{DIAL's}\\
\textbf{HPC} & \textbf{Throughput} & \textbf{Throughput}\\
\textbf{Applications} & \textbf{(MB/s)} & \textbf{(MB/s)}\\
\midrule
VPIC-IO (1D array write)	& 327.2	& 321.6\\
\midrule
VPIC-IO (2D array write)	& 319.4	& 317.7\\
\midrule
VPIC-IO (3D array write)	& 331.1	& 326.1\\
\midrule
BDCATS-IO (partial read)	& 441.3	& 436.1\\
\midrule
BDCATS-IO (strided read)	& 455.5	& 455.5\\
\midrule
BDCATS-IO (full read)	    & 463.3	& 463.1\\
\bottomrule

\end{tabular}
\label{tab:hpc_app}
\end{table}

\begin{figure}[htpb]
    \centering
    \includegraphics[width=0.48\textwidth]{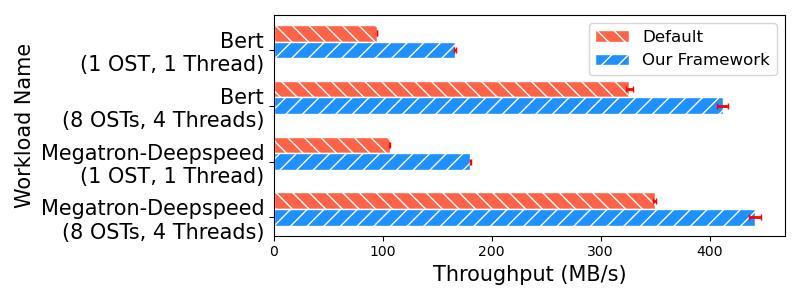}
    \caption{Deep learning application executions.}
    \label{fig:dlio_test}
    \vspace{-1em}
\end{figure}

We then evaluated two distinct deep learning I/O kernels: one representing the BERT natural language processing model and the other a Deepspeed version of NVIDIA’s Megatron language model. These I/O kernels were evaluated using the deep learning I/O benchmark (DLIO) across different utilization of object storage targets (OSTs) and varying numbers of threads. The evaluations in Figure~\ref{fig:dlio_test} revealed a notable improvement over the default configurations, with enhancements reaching as high as \textbf{1.75 times the baseline performance}. This finding highlights the framework's capacity for effectively tuning real-world applications.

\subsection{DIAL Execution Overhead}
DIAL operates independently on each client, with a design focused on efficiency in both storage and resource usage. It requires minimal storage space, maintaining only two snapshots in the client machine's main memory at any time. Additionally, because DIAL's I/O operations are limited to collecting statistics from the local file system without directly interacting with the Lustre shared file system, it avoids adding extra I/O overhead to the PFS.
Table~\ref{tab:overhead} details the average time taken for key operations.
For the evaluations, we used a very short probing interval of 0.5 seconds to probe the local file system, which was made possible by the use of local-only metrics and autonomous execution. The probing interval can be adjusted, allowing users to customize the framework's operation to suit their specific needs.

\setlength{\belowdisplayskip}{-10pt} 
\begin{table}[htbp]
\caption{DIAL Overheads per OSC Interface}
\footnotesize
\centering
\begin{tabular}{cccc}

\toprule
 & \textbf{Snapshot} &  & \textbf{End-to-end}\\
\textbf{Operation} & \textbf{Creation} & \textbf{Inference} & \textbf{Tuning}\\
\textbf{Type} & \textbf{Time (ms)} & \textbf{Time (ms)} & \textbf{Time (ms)}\\
\midrule

Read & 0.33 & 10.06 & 24.64\\
\midrule

Write & 0.85 & 13.51 & 28.82\\
\bottomrule

\end{tabular}
\label{tab:overhead}
\end{table}

\section{Related Work}
\label{sec:related}
Auto-tuning systems have been widely studied and proven complicated to explore the large optimization space.
Recent work turned to modeling HPC environment and employing black-box optimizer. Cao et al.~\cite{cao2018towards} carried out a comprehensive comparative study across multiple traditional file systems, applying five distinct auto-tuning strategies to assess the efficacy of the black-box method. SAPPHIRE~\cite{lyu2020sapphire} utilizes Bayesian Optimization to recommend the best configurations. Behzad et al.~\cite{behzad2019optimizing} implemented nonlinear regression models to model I/Os and employed genetic algorithms to explore the configuration space. 
CAPES~\cite{li2017capes} leveraged deep reinforcement learning learning to adaptively learn tuning strategies. 
However, the high overheads of accurately identifying application I/O patterns and global HPC environment make them limited in offering adaptive, real-time tuning at scale.

\section{Conclusion and Future Plan}
\label{sec:conclude}
In this study, we propose a new autotuning framework, DIAL, that enables online parameter tuning for PFSs from the client-side. By leveraging only local metrics and effective machine learning algorithms, the framework shows its generality across a wide range of workloads and scalability in large-scale systems.
In the future, we plan to modify parallel file system source code (e.g., Lustre) to expose more runtime-tunable parameters and verify the effectiveness of DIAL in more generic scenarios. 

\section*{Acknowledgments}
We sincerely thank the anonymous reviewers for their valuable feedback. This work was supported in part by National Science Foundation (NSF) under grants CNS-2008265 and CCF-2412345.

\renewcommand{\bibfont}{\footnotesize} 
\bibliographystyle{IEEEtran}
{
    \bibliography{bib}
}
\end{document}